\documentclass[prl,aps,floatfix,showpacs,nofootinbib,twocolumn,letterpaper]{revtex4-1}
\usepackage{epsfig} 
\usepackage{amssymb} 
\usepackage{amsmath} 
\usepackage{amsfonts} 
\usepackage{color, graphicx} 
\usepackage{bm}
\newcommand{\ssec}[1]{\emph{#1}.---}

\def\be{\begin{equation}} 
\def\ee{\end{equation}}

\begin{document} 

\title{Neutron width statistics in a realistic resonance-reaction model}
\author{ P.~Fanto$^{1}$, G.~F.~Bertsch$^{2}$, and Y.~Alhassid$^{1}$}
\affiliation{$^{1}$Center for Theoretical Physics, Sloane Physics Laboratory, Yale University, New Haven, CT 06520 \\
$^{2}$Department of Physics and Institute for Nuclear Theory, 
Box 351560\\ University of Washington, Seattle, WA 98195\\
}
\date{\today} 
 
\begin{abstract} A recent experiment on $s$-wave neutron scattering from $^{192,194,196}$Pt found that the reduced neutron width distributions deviate significantly from the expected Porter-Thomas distribution (PTD), and several explanations have been proposed within the statistical model of compound nucleus reactions. Here, we study the statistics of reduced neutron widths in the reaction $n+ ^{194}$Pt within a model that combines the standard statistical model
 with a realistic treatment of the neutron channel.  We find that, if the correct secular energy dependence of the average neutron widths is used, then the reduced neutron width distribution is in excellent agreement with the PTD for a reasonable range of the neutron-nucleus coupling strength and depth of the neutron channel potential. Within our parameter range, there can be a near-threshold bound or virtual state of the neutron channel potential that modifies the energy dependence of the average width from the $\sqrt{E}$ dependence, commonly assumed in experimental analysis, in agreement with the proposal of H.~A. Weidenm\"uller~\cite{Wei10}.
 In these cases, the reduced neutron width distributions extracted using the $\sqrt{E}$ dependence are significantly broader than the PTD.  We identify a relatively narrow range of parameters where this effect is significant.
\end{abstract} 
 
 
\maketitle 

\ssec{Introduction} The statistical model of compound nucleus (CN) reactions predicts that reduced widths for any channel follow the Porter-Thomas
distribution (PTD)~\cite{Porter1956,Mitchell2010}, a $\chi^2$ distribution in $\nu=1$ degrees of freedom.  
Recently, an experiment on $s$-wave neutron scattering from $^{192,194,196}$Pt found a much broader distribution of the reduced neutron widths~\cite{Koehler2010}.
Several explanations have been proposed for this deviation from the PTD within the statistical model, but none has fully resolved the issue.  

In Ref.~\cite{Wei10}, it was argued that the secular energy dependence of the average neutron widths can deviate from the usually assumed $\sqrt{E}$ form for Pt isotopes because of a near-threshold bound or virtual state of the neutron channel potential.
The authors of  Ref.~\cite{Koehler2010} showed that using the modified normalization proposed in Ref.~\cite{Wei10} [see (\ref{analytic}) below] to extract the reduced widths did not improve the agreement between their data and the PTD~\cite{comment}.  However, their procedure for determining the resonances might not hold in the presence of a state very close to threshold~\cite{Wei10, comment}, so the possible existence of this state is still an open question.
 
Other work has attempted to explain the experimental results through the non-statistical interactions between the CN states due to coupling to the neutron channel.
It has been shown that the imaginary non-statistical interaction 
can cause deviation from the PTD even for fairly weak coupling \cite{Celardo2011, Fyodorov2015}.  
However, it is not clear how strong this effect would be in Pt isotopes.  
In Ref.~\cite{Volya2015}, it was proposed that the real shift due to off-shell coupling to the neutron channel perturbs the GOE near threshold.  
However, it was subsequently proven \cite{Bogomolny2017} that in the model of Ref.~\cite{Volya2015} the PTD would hold locally in the resonance spectrum.
Many-body correlations beyond the statistical model have also been studied \cite{Volya2008}.

However, no study has incorporated all the relevant physics of the statistical model. 
Importantly, near threshold, the real and imaginary terms have a strong energy dependence that has been neglected in all prior numerical and analytical work \cite{Celardo2011,Fyodorov2015,Volya2015,Bogomolny2017,Volya2008}.
Moreover, no study has used realistic parameters for neutron scattering from Pt isotopes.  
For these reasons, prior work has not fully settled the question of whether PTD violation within the statistical model could occur for this reaction.  This problem is of considerable importance because the statistical model is widely used in reaction calculations.

Here, we study neutron scattering off $^{194}$Pt within a reaction model that combines a
realistic treatment of the neutron channel with the usual
description of the internal CN states by the Gaussian
orthogonal ensemble (GOE) of random-matrix theory~\cite{Mitchell2010}.  
Our model enables us to study average neutron widths, the reduced width distribution, and the elastic and capture cross sections within the same framework.  
We start with a baseline physical parameter set for the model taken from the literature.  We then vary the parameter set to produce the conditions under which the proposed mechanisms for PTD violation could be operative.
Finally, we discuss the compatibility of these varied parameter sets with the scattering data.

Our main conclusion is that, within the reasonably large parameter range studied, the reduced neutron width distribution is in excellent agreement with the PTD. Thus, when described realistically, the non-statistical interactions
cannot explain the observed deviation from the PTD within the parameter range used.
Evidence of PTD violation 
may be observed only if the secular energy dependence of the average neutron width is not described correctly.  
Within our parameter range, there can be a near-threshold bound or virtual state of the neutron channel potential.  
In the presence of such a state, the energy dependence of the average neutron width differs significantly from the $\sqrt{E}$ dependence~\cite{Wei10}, and reduced width distributions extracted with the $\sqrt{E}$ assumption are significantly broader than the PTD.  
We identify measurable signatures of this state's existence.

\ssec{Hamiltonian and resonance determination}
Our model Hamiltonian matrix $\mathbf{H}$ combines a mesh representation of the neutron channel with the GOE description of the internal states.  The neutron channel mesh has spacing $\Delta r$ and radial sites $r_i = i \Delta r$, $(i = 1,...,N_n)$.  The channel Hamiltonian matrix is $\mathbf{H_{n}}_{,ij} = [2t + V(r_i)]\delta_{ij} - t \delta_{i,j+1} - t\delta_{i,j-1}$, where $t = \hbar^2/2m(\Delta r)^2$ and $V(r)$ is the channel potential.
The energies of the $N_c$ internal states follow the middle third of a GOE spectrum with average spacing $D$.  To each internal energy we add the imaginary constant $(-i/2)\Gamma_\gamma$ to account for resonance decay by gamma-ray emission.  The neutron channel couples to each internal state $\mu$ at a single site $r_e = i_e\Delta r$ with strength $v_\mu = v_0 (\Delta r)^{-1/2} s_\mu$, where $v_0$ is a coupling constant and $s_\mu$ are drawn from a normal distribution with zero average and unit variance.  The explicit $\Delta r$ dependence of $v_\mu$ is required to achieve a fixed $v_0$ in the continuum limit $\Delta r\to 0$.
All results shown below were calculated using $(\Delta r, N_n, N_c) = (0.01\text{ fm}, 1500, 360)$.

We find the complex wavenumbers $k_r$ that correspond to the neutron resonances by solving the Schr\"odinger equation $\mathbf{H}\,\vec u =E \vec u$ ($\vec u$ is a column vector with $N_n+N_c$ components) with the appropriate boundary conditions for the neutron wavefunction $u(r)$.
We impose $u(0) = 0$ for the wavefunction to be regular at the origin.
A resonance is a pole of the $S$ matrix corresponding asymptotically to a purely outgoing wave, 
i.e.~$u(r) \rightarrow B(k) e^{ik r}$ for large $r$.  
For sufficiently large $N_n$, this condition yields 
$u(N_n + 1) = u(N_n) e^{ik\Delta r}$. 
We obtain the nonlinear eigenvalue problem 
\be\label{evp} 
\mathbf{M}(k) \vec{u} =\left[\mathbf{H} - t e^{ik\Delta r}\mathbf{C} - E \right]\vec{u} = 0 
\ee
where $\mathbf{C}_{ij} = \delta_{i,j} \delta_{i,N_n}$.  We solve (\ref{evp}) iteratively to find the resonances $k_r$, adapting a method from Ref.~\cite{by13}.  
The resonance energies $E_r$ and total widths $\Gamma_r$ are determined from  $\hbar^2 k_r^2/2m=E_r -( i/2 )\Gamma_r $.  
The partial neutron widths $\Gamma_{n,r}$ are then given by $\Gamma_{n,r} = \Gamma_{r} - \Gamma_\gamma$.  
Elastic and capture cross sections are calculated from the elastic scattering amplitude, which is determined using the boundary conditions of a scattering wave.  Further details and the relevant computer codes are provided in the Supplementary Material~\cite{supp_material}.

\begin{figure}[h!]
\includegraphics[width=0.5\textwidth]{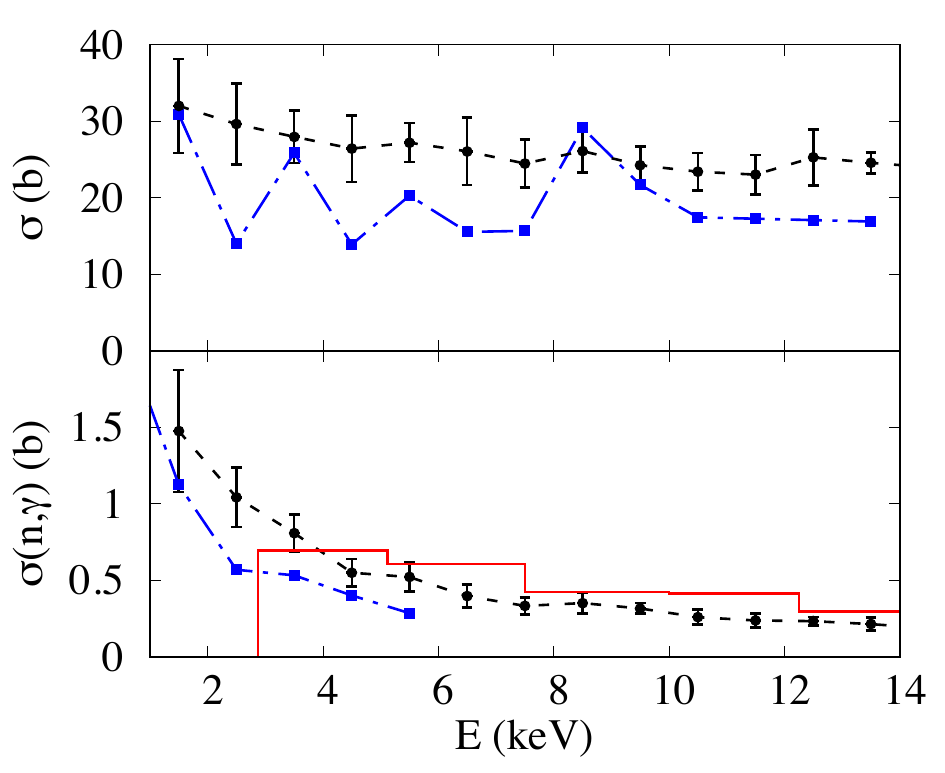}
\caption{\label{baseline} 
Elastic scattering (top panel) and capture (bottom panel) cross sections.
Our baseline calculations, averaged over 1 keV bins (black circles joined by dashed line), are compared with cross sections from the JEFF-3.2 library~\cite{jeff}, averaged over the same bins (blue squares joined by dashed-dotted line).  
Error bars indicate standard deviations from 10 realizations of the GOE.  
The red histogram shows experimental average capture cross sections~\cite{ko13}.
}
\end{figure}

\ssec{Application to $n+ ^{194}${\rm Pt}} 
We determine a baseline parameter set as follows.
We take a Woods-Saxon potential in the neutron channel with parameters $V_0 = -44.54$ MeV and $(r_0,a_0) 
= (1.27,0.67)$ fm [see Eqs.~(2-181) and (2-182) of Ref.~\cite{Bohr1969}]. 
The mean resonance spacing $D = 82$ eV and the total gamma decay width
$\Gamma_\gamma = 72$ meV are taken from the RIPL-3 database \cite{ripl}.  
We choose a coupling strength of $v_0 = 11$ keV-fm$^{1/2}$ to reproduce roughly
 the RIPL-3 neutron strength function $S_0\sqrt{E_n} = \bar{\Gamma}_n/D$ at neutron energy of $E_n = 8$ keV (see Table~\ref{T1}).   

Fig.~\ref{baseline} shows the elastic and capture cross sections for the baseline model 
averaged over neutron energy in bins of 1 keV width.  
We also show elastic and capture cross sections from the JEFF-3.2 library~\cite{jeff}, which are based on the reaction code TALYS~\cite{ko12}, averaged over the same energy bins.  
The histogram in the bottom panel of Fig.~\ref{baseline} shows experimental energy-averaged capture cross sections~\cite{ko13}.  
Overall, the agreement with other calculations and experiment is sufficiently close to take the baseline parameter set as our starting point. 

\begin{figure}[h!]
\includegraphics[width=.5\textwidth]{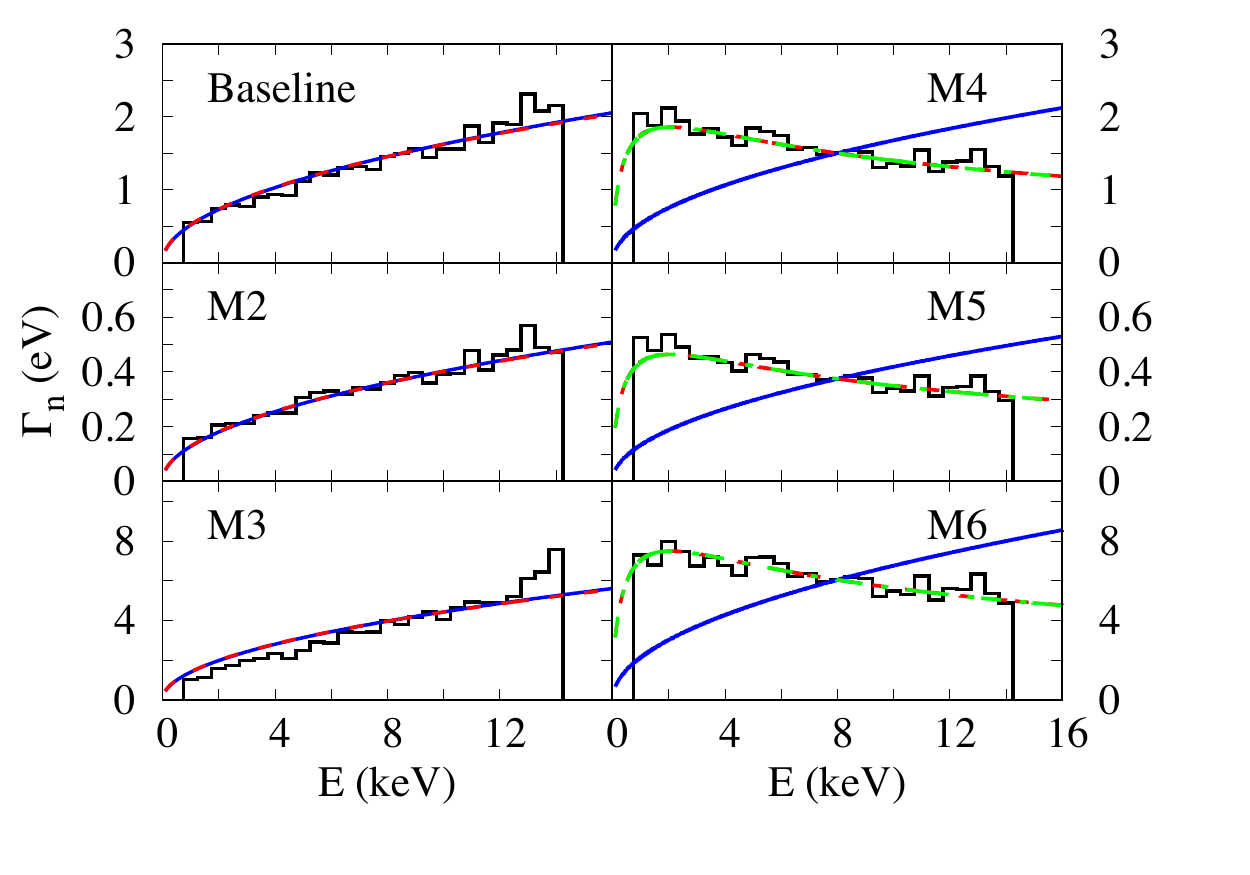}
\caption{\label{gnbar} 
Comparison of $\bar{\Gamma}_n(E)$ calculated for the different models of Table \ref{T1} (histograms) with $\sqrt{E}$ (solid blue lines), the neutron probability density  $u^2_E(r_e)$ (red dashed lines), and the formula in Eq.~(\ref{analytic})~\cite{Wei10} (green dashed-dotted lines).  
Functions are normalized to match the model calculations at $E = 8$ keV.}
\end{figure}

\ssec{Reduced neutron width statistics} The reduced neutron width $\gamma_{n,r}$ is defined by
\be\label{reduced}
\gamma_{n,r} = {\Gamma_{n,r} / \bar{\Gamma}_n(E_r)} \;,
\ee
where $\bar{\Gamma}_n(E)$ is the average width that varies smoothly with the neutron energy $E$.
Fig.~\ref{gnbar} shows the average widths calculated for various parameter sets.  
In each case, the data was computed for 100 GOE realizations, from each of which we take as data 160 resonances from the middle of our model resonance spectrum.  The real parts of these resonance energies fall mostly in the interval $E=1-14$ keV, which covers the bulk of the experimental range of Ref.~\cite{Koehler2010}.  
For the baseline model, the histogram compares well with the $\sqrt{E}$ dependence. 
The probability density of the neutron scattering wavefunction~\cite{normalization} at the interaction point, $u^2_E(r_e)$, is also shown in Fig.~\ref{gnbar} and, for the baseline model, is hardly distinguishable from the $\sqrt{E}$ curve, in agreement with the statistical model prediction~\cite{Wei10, Mahaux1969}.

\begin{figure}[h!]
\centering
\includegraphics[width=0.5\textwidth]{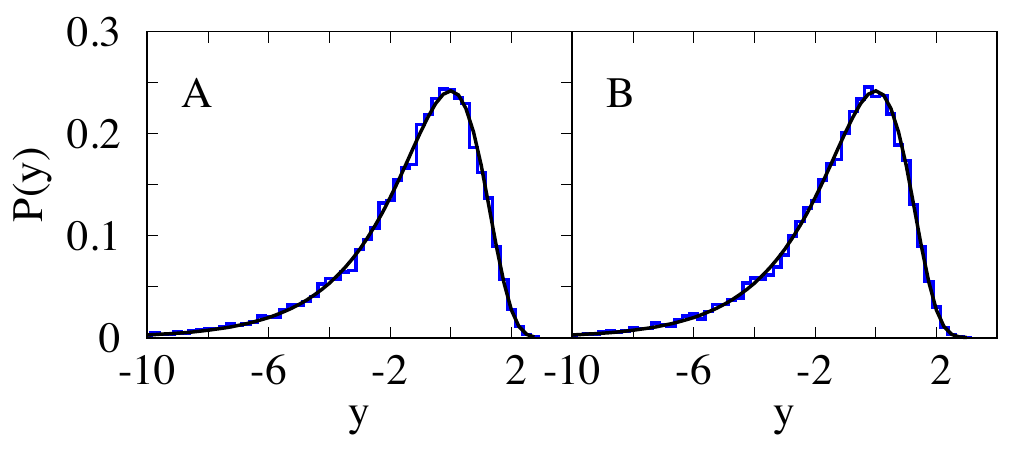}
\caption{\label{stat-44} 
The histograms describe the distributions of the logarithm of the normalized reduced width for the baseline model.  The reduced widths are calculated from Eq.~(\ref{reduced}).  Reduction A (left panel) uses $\bar \Gamma_n(E)$ from the model, while reduction B (right panel) uses $\bar \Gamma_n(E) \propto \sqrt{E}$.
The solid lines are the PTD.}
\end{figure}

Next, we determine the reduced widths and compare their distribution with the PTD.
We consider the distributions extracted using the average widths calculated from the model, which we call reduction A, as well as those extracted using the $\bar \Gamma_n(E) \propto \sqrt{E}$ ansatz, which we call reduction B.  
Fig.~\ref{stat-44} shows as histograms the calculated probability distributions of the logarithm $y = \ln x$ of the normalized 
reduced widths $x = \gamma_n/\langle \gamma_n \rangle$ for the baseline model.
For both reductions A and B, we find excellent agreement with the PTD for $y$
\be\label{ptd}
\mathcal{P}(y) = x \mathcal{P}_{\rm PT}(x)  = \sqrt{\frac{x}{2 \pi}}e^{-x/2}  \;.
\ee
 For a quantitative comparison, we compute the reduced chi-squared value  $\chi^2_r$, using $\chi^2_r \approx 1$ as a criterion for a good fit \cite{chi-squared}.
The baseline model yields $\chi_r^2 \approx 1$ for both reductions A and B (see Table \ref{T1}).

\begin{table}[h!]
\begin{tabular}{| c | ccc | ccc  |}
\hline
Model & baseline & M2  & M3 & M4 & M5 & M6\\ \hline
$V_0$ (MeV) & \multicolumn{3}{c|}{-44.54} & \multicolumn{3}{c|}{-41.15} \\ 
$v_0$ (keV-fm$^{1/2}$) &  11.0 & 5.5  & 22.0 & 1.6 & 0.8 & 3.2 \\ \hline
$S_0 \cdot 10^{4}$ (eV$^{-1/2}$) & 2.0 & 0.5 & 5.4 & 2.0 & 0.5 & 8.2 \\
$ \bar\sigma_{el}$ (b) & 30. & 19.0 & 23. & 279. & 288.  & 249.  \\
$ \bar\sigma_\gamma$ (b) & 0.44 & 0.32 & 0.50 & 0.47 & 0.39 & 0.53  \\
$\chi_r^2 \,\,\,$ PTD A& 0.9 & 1.0 & 1.1 & 0.9 & 1.0 &  1.4  \\
$\chi_r^2 \,\,\,$ PTD B & 1.0 & 1.0 &  1.3 & 5.8 & 6.0 & 6.1 \\
$\nu_{\rm fit}\,\,\,$ A & 1.0  & 1.0 & 0.98 & 1.0 & 1.0 & 0.98 \\
$\chi_r^2\,\,\,$ fit A & 0.9 & 1.0 & 1.0  & 0.9 & 1.0 & 1.3 \\
$\nu_{\rm fit}\,\,\,$ B &  1.0 & 1.0 & 0.97 & 0.92 & 0.92 & 0.92 \\
$\chi_{r}^2\,\,\,$ fit B & 1.0 & 1.1 & 1.1 & 3.4 & 3.8 & 3.7 \\
\hline
\end{tabular}
 \caption{\label{T1}  
Calculated resonance properties of the $n+^{194}$Pt reaction for various parameter sets.  
The neutron strength function parameter $S_0 = (\bar \Gamma_n /D)/\sqrt{ E}$ and average
elastic scattering cross section $\bar\sigma_{el}$ are evaluated
at $E= 8$ keV.  The RIPL-3 strength function parameter is $2\cdot 10^{-4}$ eV$^{-1/2}$ \cite{ripl}.  The capture cross section $\bar \sigma_\gamma$ is the average over the interval 5-7.5 keV corresponding to the measured value of 0.6 b~\cite{ko13}.   
Reductions A and B are as described in the caption to Fig.~\ref{stat-44}.
The row labeled $\chi_r^2$ PTD contains the chi-squared results comparing the reduced width distributions to the PTD.
The values $\nu_{\rm fit}$ and $\chi^2_r$ fit refer to the maximum-likelihood fit  to Eq.~(\ref{nu}).
}
 \end{table}

\ssec{Parameter variation}  Here we vary the parameters $v_0$ and $V_0$ to investigate proposed explanations for PTD violation.
 First, we vary the coupling strength $v_0$ by a factor
of two smaller or larger than the baseline value, keeping $V_0$ fixed at its baseline value.
These sets are labeled, respectively, by M2 and M3 in Table \ref{T1}.  As shown in Table \ref{T1}, the average elastic scattering cross section at $E=8$ keV varies only in the range 19--30 b, and the average capture cross section in the interval 5--7.5 keV varies by a similar fractional amount.  The reduced width distributions from reductions A and B are nearly identical to the corresponding baseline distributions in Fig.~\ref{stat-44}.
The $\chi^2_r$ values for the PTD  are all close to 1, indicating good agreement with the PTD.  
In the strong coupling case M3, the average width shown in Fig.~\ref{gnbar} deviates somewhat from the expected $\sqrt{E}$ dependence.  This is a numerical effect due to the finite bandwidth of internal states~\cite{supp_material}.

Next, we vary $V_0$ to investigate the effect of a near-threshold bound or virtual state in the neutron channel.  With our baseline potential, there is a bound $4s$ neutron level at energy $\approx -0.7$ MeV \cite{bound}.  Changing $V_0$ to $-41.15$ MeV results in a weakly bound state with energy $E_0 \approx -2$ keV.
This change in $V_0$ is sufficiently moderate to justify its inclusion in our parameter set~\cite{V0}.
We adjust $v_0$ in model M4 to reproduce the RIPL-3 strength function parameter $S_0$ and vary $v_0$ by a factor of two smaller or larger for models M5 and M6, respectively.
The average capture cross sections for models M4--M6, shown in Table \ref{T1}, are only slightly larger than those of the baseline model.
However, the elastic cross sections are much larger than the baseline values.  Thus,  experimental elastic cross sections could be used to narrow the parameter values of our model.  
Unfortunately, we know of no published experimental elastic cross sections for this reaction.  

As shown in Fig.~\ref{gnbar}, the average neutron widths for models M4--M6 have an energy dependence that differs significantly from $\sqrt{E}$.  
 However, the quantity $u^2_E(r_e)$ remains an excellent estimator of the correct energy dependence of the average widths. 
An analytic expression was derived in Ref.~\cite{Wei10}
for a near-threshold bound or virtual state with energy $E_0$ ($E_0 < 0$)  
\be
\label{analytic}
u^2_E(r_e) \propto {\sqrt{E}\over E + |E_0|}\,.
\ee
Using $E_0 \approx -2$ keV from our model in Eq.~(\ref{analytic}), we find excellent agreement with
both $u^2_E(r_e)$ and the average widths (see Fig.~\ref{gnbar}).  

\begin{figure}[h!]
\includegraphics[width=0.5\textwidth]{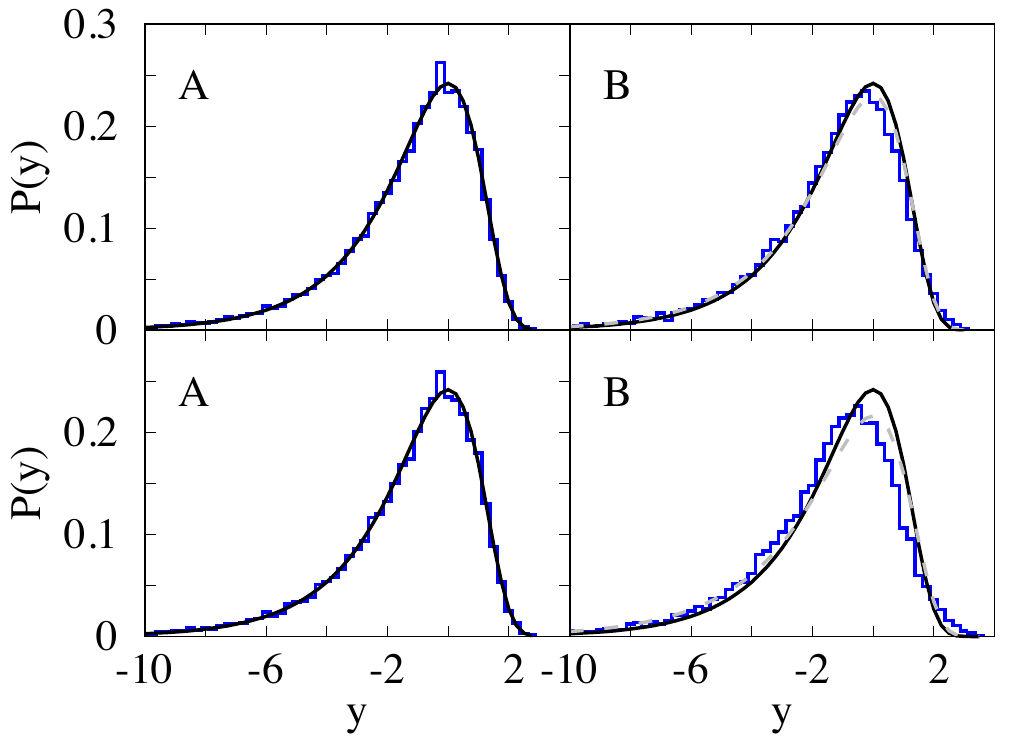}
\caption{\label{stat-40}  As in Fig.~\ref{stat-44} but for model M4 (top panels), and for the model with $E_0 \approx 0$ (bottom panels). 
$\chi^2$ distributions in $\nu  = \nu_{\rm fit}$ degrees of freedom are shown by the dashed gray lines.  See text for details.
}
\end{figure}

The reduced width distributions for model M4 are shown in the upper panels of Fig.~\ref{stat-40} (similar results are obtained for models M5 and M6). 
The distributions extracted with the calculated $\bar \Gamma_n(E)$ (reduction A) are well described by the PTD,  
as is confirmed by the $\chi_r^2$ values in Table \ref{T1}.  In contrast, the distributions obtained using the $\sqrt{E}$ dependence (reduction B) are noticeably broader than the PTD, and the $\chi_r^2$ values for this reduction are significantly larger than 1. 

As we make the neutron potential slightly less attractive, the weakly bound state becomes a virtual state whose energy $E_0$ is also negative but on the second Riemann sheet~\cite{virtual}. For example, when $V_0=-40.85$ MeV, we have a virtual state with $E_0\approx-2$ keV.   According to (\ref{analytic}), the maximal deviation of the average width from $\sqrt{E}$ occurs for $E_0=0$. We then expect to see the maximal deviation from a PTD in reduction B.  In our model, this occurs for $V_0 = -41$ MeV.  The reduced width distributions for this case are shown in the lower panels of Fig.~\ref{stat-40}. For reduction B, we observe an even stronger deviation from the PTD, as expected.

Finally, for all the parameter sets considered, we made a maximum-likelihood fit of the calculated distributions to a $\chi^2$ distribution in $\nu$ degrees of freedom
\be
\label{nu}
\mathcal{P}(x|\nu) = {\nu (\nu x)^{\nu/2-1}\over 2^{\nu/2} \Gamma({\nu/2})} e^{-\nu x /2} \;.
\ee
 More specifically, we find the value $\nu_{\rm fit}$ that maximizes the likelihood function $L(\nu) = \prod_i\mathcal{P}(x_i|\nu)$, 
where $x_i$ are the reduced width data values.  The PTD is recovered for $\nu = 1$.  As shown in Table \ref{T1}, for reduction A, all models reproduce the PTD. 
Moreover, for reduction B, models M1-M3 also reproduced the PTD.  However, for models M4-M6 and for reduction B, we obtain $\nu_{\rm fit} = 0.92$ for all cases, and the $\chi_r^2$ values are significantly larger than 1.

\ssec{Conclusion} 
We have studied the statistics of neutron resonance widths in the $n + ^{194}$Pt reaction within a model that combines a realistic treatment of the neutron channel with the GOE description of the internal states.  Our model is the first to incorporate all aspects of the statistical model for a single-channel reaction.  
Our main conclusion is that the PTD describes well the distribution of reduced neutron widths (\ref{reduced}) for a reasonably large parameter range around baseline values taken from the literature. 
Our results indicate that non-statistical interactions do not explain the experimentally observed PTD violation.  
These interactions may be more important in other systems, where the coupling between the channels and the internal states is stronger.

Apparent PTD violation may be observed only if the secular energy dependence of the average neutron width is not described correctly.  Within our parameter range, this can happen in the presence of a near-threshold bound or virtual state of the neutron channel potential.  
In this case, the energy dependence of the average width differs significantly from $\sqrt{E}$, 
and the distributions of reduced widths extracted with the usual $\sqrt{E}$ ansatz are broader than the PTD.
However, significant deviations from the $\sqrt{E}$ behavior require that the magnitude $|E_0|$ of the energy of this near-threshold state 
be no more than a few keV for $^{192,194,196}$Pt.  
Moreover, as stated above, the authors of Ref.~\cite{Koehler2010} showed that using the form (\ref{analytic}) did not improve their data's agreement with the PTD~\cite{comment}.
However, a state so close to threshold might undermine the experimental resonance determination procedure \cite{Wei10,comment}.
We have found that the magnitude and shape of the elastic neutron cross section are strongly affected by a near-threshold state in the neutron channel potential (see Table \ref{T1}).
Therefore, experimental measurements of the elastic cross section would be useful in determining the possible existence of  such a near-threshold state. 

\ssec{Acknowledgments} This work was supported in part by the U.S. DOE grant
Nos.~DE-FG02-00ER411132 and DE-FG02-91ER40608, and by the DOE NNSA Stewardship Science Graduate Fellowship under cooperative agreement No.~DE-NA0002135.  
We would like to thank  H.~A. Weidenm\"uller for useful discussions. PF and YA acknowledge the hospitality of the Institute for Nuclear Theory at the University of Washington, where part of this work was completed during the program INT-17-1a, ``Toward Predictive Theories of Nuclear Reactions Across the Isotopic Chart.''  This work was supported by the HPC facilities operated by, and the staff of, the Yale Center for Research Computing.


\begin{thebibliography}{100}
\bibitem{bound} Like a resonance, a bound state is a pole of the $S$ matrix.  We find the bound state energies of the Woods-Saxon potential using the same method by which we determine the resonance energies for our model Hamiltonian \cite{supp_material}.
\end{thebibliography}
\end{document}